\UseRawInputEncoding
\documentclass[%
 reprint,
 amsmath,amssymb,
 aps,
]{revtex4-2}

\usepackage{graphicx}
\usepackage{dcolumn}
\usepackage{bm}
\usepackage{color}

\begin{document}

\preprint{APS/123-QED}

\title{Effect of nitrogen introduced at the SiC/SiO$_2$ interface and SiC side on the electronic states by first-principles calculation}

\author{Keita Tachiki}
\email{tachiki@semicon.kuee.kyoto-u.ac.jp}
\affiliation{
Laboratory for Materials and Structures,
Institute of Innovative Research,
Tokyo Institute of Technology,
Yokohama 226-8503,
Japan;
}

\affiliation{
Quemix Inc.,
Taiyo Life Nihombashi Building
2-11-2,
Nihombashi Chuo-ku, 
Tokyo 103-0027,
Japan;
}
\affiliation{[Presently at] the Department of Electronic Science and Engineering, Kyoto University, Nishikyo, Kyoto, 615-8510, Japan}

\author{Yusuke Nishiya}%
\affiliation{
Laboratory for Materials and Structures,
Institute of Innovative Research,
Tokyo Institute of Technology,
Yokohama 226-8503,
Japan;
}
\affiliation{
Quemix Inc.,
Taiyo Life Nihombashi Building
2-11-2,
Nihombashi Chuo-ku, 
Tokyo 103-0027,
Japan
}

\author{Jun-Ichi Iwata}
\affiliation{
Laboratory for Materials and Structures,
Institute of Innovative Research,
Tokyo Institute of Technology,
Yokohama 226-8503,
Japan;
}
\affiliation{
Quemix Inc.,
Taiyo Life Nihombashi Building
2-11-2,
Nihombashi Chuo-ku, 
Tokyo 103-0027,
Japan
}

\author{Yu-ichiro Matsushita}
\affiliation{
Laboratory for Materials and Structures,
Institute of Innovative Research,
Tokyo Institute of Technology,
Yokohama 226-8503,
Japan;
}
\affiliation{
Quemix Inc.,
Taiyo Life Nihombashi Building
2-11-2,
Nihombashi Chuo-ku, 
Tokyo 103-0027,
Japan;
}
\affiliation{
Quantum Material and Applications Research Center,
National Institutes for Quantum Science and Technology,
2-12-1, Ookayama, Meguro-ku, Tokyo 152-8552, Japan
}




\date{\today}

\begin{abstract}
In this study, using first-principles calculations, we investigate the behavior of electrons at the SiC/SiO$_2$ interface when nitrogen is introduced as a dopant within a few nm of the SiC surface. When a highly doped nitrogen layer (5$\times$10$^{19}$ cm$^{-3}$) is introduced within a few nm of the SiC(11$\bar{2}$0) surface, the electronic state is not significantly affected if the doping region is less than 4 nm. However, if the doping region exceeds 4 nm, the effect of quantum confinement decreases, which increases the electron density induced in the inversion layer. As for the wave function, even when an electric field is applied, the peak shifts toward the direction in which the electrons are pulled away from the interface. This reduces the effect of electron scattering at the interface and improves electron mobility. 
\end{abstract}

\maketitle


\section{\label{sec:Introduction} Introduction}
SiC has attracted considerable attention as a material for realizing ultra-low-loss switching devices owing to its excellent physical properties, and its practical applications are already being investigated \cite{power1, power2, power3, Kimoto}. Theoretically, by replacing SiC-MOSFETs, the conduction energy loss can be reduced to 1/500 compared to that of Si. However, the actual switching energy loss of SiC-MOSFETs is far from the theoretical value and instead reaches a maximum of approximately 1/10. This is due to the high-density defects at the SiC/SiO$_2$ interface \cite{mobility1, mobility2, vilak, sak, ortiz, uhne, mobility4, mobility5, sometani, mobility6, 79, hirai, noguchi, Kobayashi1, Kobayashi2, matsushita4, chokawa}. Experimental studies have found that high-temperature annealing in nitric oxide after the formation of gate oxide is effective in reducing  these high-density interface defects \cite{NO1, NO2, NO3, NO4, NO5}; this technique has been widely used as a passivation method. However, the obtained channel mobility is still not sufficiently high, and it is unclear why nitridation \cite{NO1, NO2, NO3, NO4, NO5, N21, toshiba, tachiki} improves channel mobility.

Secondary ion mass spectrometry (SIMS) and X-ray photoelectron spectroscopy (XPS) have shown that nitridation introduces a large amount ($\sim$1$\times$10$^{21}$ cm$^{-3}$) of nitrogen at the SiC/SiO$_2$ interface \cite{SIMS, XPS}; however, how this affects interface defects is unclear. N introduced at the SiC/SiO$_2$ interface is hypothesized to terminate the dangling bonds. However, it was recently reported that N diffuses not only into the interface but also into SiC by nitridation, though there is very limited research on how this affects switching devices \cite{NSiC1, NSiC2, NSiC3, NSiC4}. In this study, using first-principles calculations, we aimed to clarify how the electronic structure is affected by N near the SiC/SiO$_2$ interface when N is densely introduced as a dopant. We also calculated the changes in the wave function and local density of states (LDOS) when an external electric field was applied, assuming the on-state of an actual MOSFET.

\section{\label{sec:Computation details}Computation details}

The calculations in this study were performed using a real-space density-functional theory (RSDFT) program code and Quloud-RSDFT tools \cite{RSDFT,Quloud} on the basis of the DFT \cite{DFT1, DFT2}. Generalized gradient approximation (GGA), specifically Perdew—Burke--Ernzerhof \cite{PBE}, was used as the exchange and correlation energies. A norm-conserving pseudopotential \cite{Normconserving} of the Troullier--Martins type \cite{TM} was used to describe the interaction between ions and valence electrons. The lattice constants were optimized using perfectly crystalline 4H-SiC. Consequently, the lattice constants obtained were a = 3.094 \AA~ and c = 10.104 \AA. The experimental lattice constants a = 3.08 \AA~ and c = 10.08 \AA~ were reproduced with an error of less than 0.7\% \cite{lattice}. 

To reproduce the SiC(11${\bar 2}$0)/SiO$_2$ interface structure, a slab model of 4H-SiC(11${\bar 2}$0) was used in this calculation. In the slab model, a 1 × 1 structure was adopted in the in-plane direction; moreover, 12 atomic layers (37 \AA) in the thickness direction and a vacuum layer thickness of 20 \AA were adopted. Hence, the total number of atoms in the simulation cell was 216, wherein the total number of Si and C atoms was 200 and the number of hydrogen atoms was 16 (a = 5.36 \AA, b = 61.9 \AA, c = 10.1 \AA). Here, the thickness of the vacuum layer was set such that the interaction between adjacent SiC slabs was sufficiently small, that is, less than 5.1e-4 eV/A$^2$. The thickness of the 4H-SiC slab was set such that it was sufficiently thick to be considered as a bulk in the deep region of the slab. Specifically, the band-gap in the slab calculation was 2.36 eV, which is sufficiently close to the band-gap of 2.27 eV in the bulk calculation. 

To model the interface, both surfaces of the slab model had hydrogen terminals. The hydrogen atoms on one surface mimicked a semi-infinite SiC substrate, which compensated for the dangling bonds appearing on the surface. In contrast, the hydrogen atoms on the other surface mimicked the connection to the SiO$_2$ film because the offset of the conduction band ($E_{\rm c}$) for SiC/SiO$_2$ was approximately 2.7 eV, which was sufficiently high and enabled approximating the SiC/SiO$_2$ interface to a hydrogen-terminated SiC surface. 

\begin{figure*}
\includegraphics{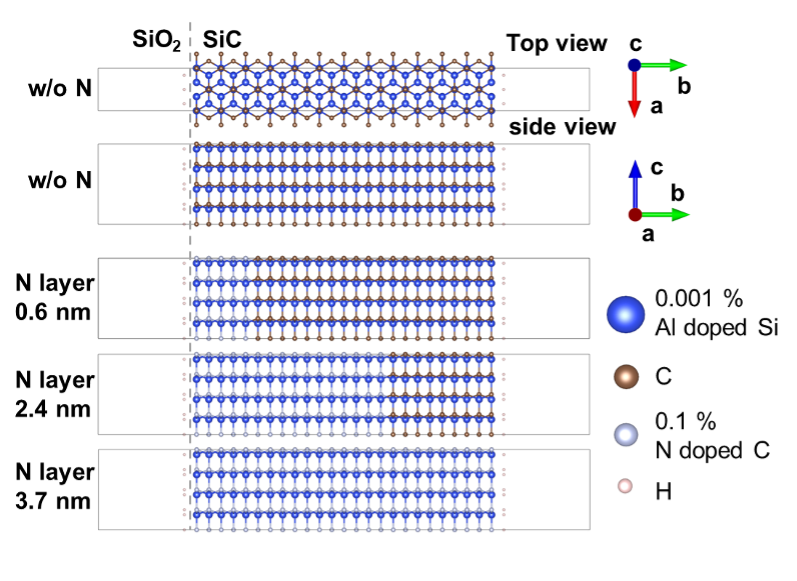}
\caption{\label{fig:fig1}(a) Slab of 4H-SiC consisting of 216 atoms used in the calculations of this study. Si and C on the surface are terminated by H. Blue, brown, white, and pink circles indicate 0.001\% Al-doped Si, perfect C, 0.1\% N-doped C, and hydrogen, respectively. The transverse length of the slab (in direction b) is 40 \AA, and the combined thickness of the vacuum layers is 20 nm. (b) Slab of 4H-SiC consisting of 864 atoms used in the calculations of this study. Blue, brown, white, and pink circles indicate 0.001\% Al-doped Si, perfect C, N, and hydrogen, respectively.}
\end{figure*}

This study focuses on the effect of nitridation on the electronic state at the interface; here, we assume an ideal situation in which there are no defects at the interface. The prepared slab is shown in Fig. ~\ref{fig:fig1}. In this study, four slab models with different doped nitridation profiles were prepared. 
The cutoff energy was set to 146 Rydberg and the sampling k-points were set to 1 × 1 × 1 ($\Gamma$ point one-point calculation). Concerning structural optimization, the forces converged to less than 5.0e-4 Hartree/Bohr.
To investigate the doping effect of nitridation on the SiC side near the interface, which is the focus of this study, a virtual crystal approximation (VCA) was employed. Specifically, pseudopotentials of Si and C atoms were prepared for the cases of Al and N doping, respectively, assuming 0.001\% Al doping for Si atoms, and no doping and 0.1\% N doping for C atoms. 
In terms of doping density, these values correspond to approximately 5$\times$10$^{17}$ and 5$\times$10$^{19}$ cm$^{-3}$, respectively. In real devices, to prevent short-channel effects, 1$\times$10$^{17}$ cm$^{-3}$ or a higher doping concentration for Al is often used. As for the nitrogen doping density, the nitrogen dopant introduced by the epitaxial growth of SiC is generally 2$\times$10$^{19}$ cm$^{-3}$ because of the temperature dependence of the solid solution limit \cite{epi}. However, SIMS analysis of nitrided MOS interfaces showed that the N density within 3 nm of the interface was very high, approximately 1$\times$10$^{19}$ to 1$\times$10$^{21}$ cm$^{-3}$ \cite{XPS, Kimoto, tachiki, toshiba}. Hence, a value of 5$\times$10$^{19}$ cm$^{-3}$ was used, although this may be slightly high for the doping density.

In this study, to investigate the nitrogen atoms uniformly distributed throughout SiC, we used VCA, which averages out the effect of nitrogen. First, we examined the validity of VCA for the purpose of this study. To verify the validity, a comparison was made between the supercell with substitution and VCA. As a validation, we checked how the difference between the potential and vacuum level in the SiC slab changes between VCA and nitrogen substitution compared to the undoped case. In conclusion, it was found that VCA reproduced the change in SiC potential and vacuum level in the slab well. Therefore, we concluded that VCA yields reasonable results for the present study. The applied electric field was constant in all directions of the slab: 0, 5, and 10 MV/cm.

\section{\label{sec:Results and discussion}Results and discussion}

\begin{figure*}
\includegraphics{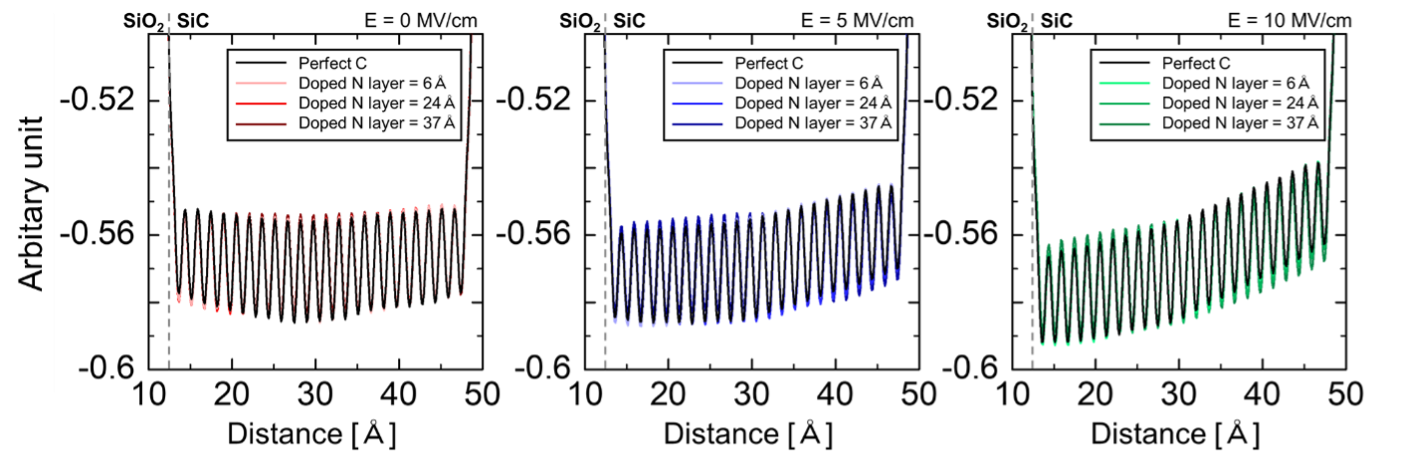}
\caption{\label{fig:fig2}Distance dependence of potential in the slabs with different N-doped regions. Applied electric fields are 0, 5, and 10 MV/cm, respectively. For clarity, averaging is performed over the atomic layer thickness period.}
\end{figure*}

Fig.~\ref{fig:fig2} shows the calculated potentials of SiC. The potential distribution was averaged over the atomic layer period in the thickness direction. In the figure, a, b, and c show the results for no applied electric field, 5 MV/cm, and 10 MV/cm, respectively. The gradient of the potential increases as the applied electric field strength increases. The electric field in the semiconductor can be roughly estimated from the potential gradient to be 1.2 MV/cm when the external electric field is 5 MV/cm, and 2.4 MV/cm when the external field is 10 MV/cm (the breakdown strength of SiC in the (11${\bar 2}$0) direction is approximately 3.3 MV/cm). Depending on the doping density, the oxide film electric field of SiC-MOSFETs in the on-state is approximately 3 MV/cm. Based on the relative permittivity of SiO$_2$ and SiC, the surface electric field of SiC at this time was considered to be approximately 1.1-1.2 MV/cm. Therefore, when the external electric field was 5 MV/cm, it is assumed to be close to the intensity of the SiC surface electric field when driving a real device.

Next, the relationship between nitrogen doping and the interface potential is discussed. Fig.~\ref{fig:fig2} shows that the potential changes with the nitrogen doping depth. When the nitrogen-doped regions are 0.6 nm and 2.4 nm, there is no significant difference, but when the region is 3.7 nm, the potential flattens (smaller gradient) in response to the electric field application compared to that without doping. Under the present calculation conditions, SiC is a p-type semiconductor with a doping density of approximately 2e17 cm$^{-3}$ without nitrogen doping, but when nitrogen is doped in the 3.7 nm region, it is considered an n-type semiconductor with a doping density of approximately 2$\times$10$^{19}$ cm$^{-3}$. The amount of band bend in the semiconductor associated with an external electric field is also consistent with the theoretical prediction that the higher the doping density, the smaller is the band bend.

In SiC-MOSFETs, in addition to interfacial roughness scattering, a large number of interfacial defects exist at the SiC/SiO$_2$ interface, which are said to be Coulomb scattering sources and reduce the carrier mobility \cite{vilak, ortiz, sak, hirai, noguchi}. Therefore, if the wave function is moved away from the interface, it is less susceptible to scattering and the carrier mobility is thought to be improved. The flattening of the potential owing to nitrogen doping is assumed to work in the direction of which the carrier wave function is moved away from the interface.

\begin{figure*}
\includegraphics{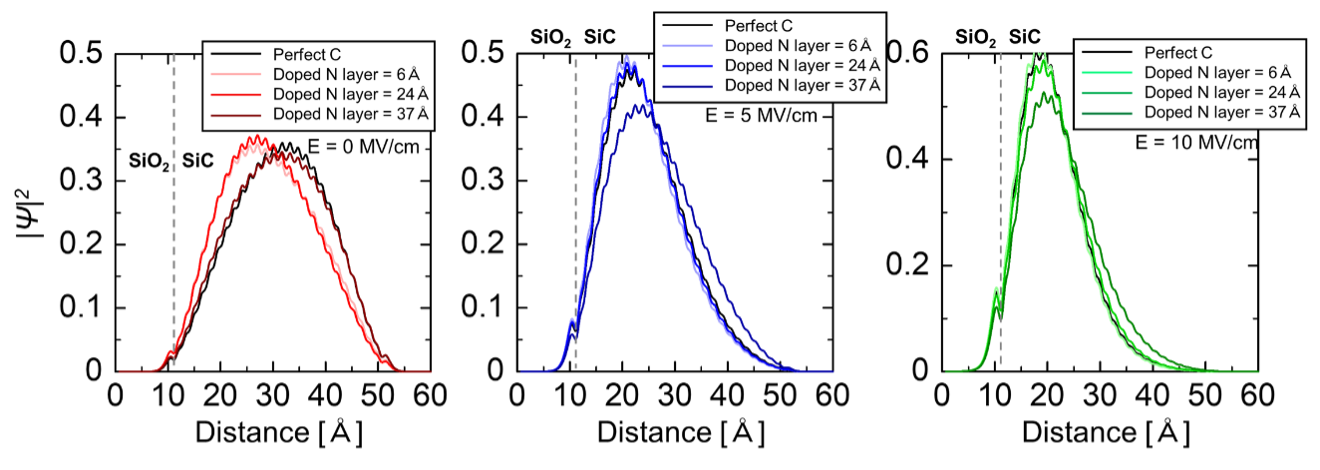}
\caption{\label{fig:fig3}LDOS of the CBM in slabs with different N-doped regions. For clarity, averaging was done over the atomic layer thickness period. Applied electric fields are 0, 5, and 10 MV/cm, respectively.}
\end{figure*}

Fig.~\ref{fig:fig3} shows the conduction band minimum (CBM) wave function. Notably, in 4H-SiC, the CBM is located at the equivalent three-fold degenerate M-point owing to the crystal symmetry; however, in this A-face, the three-fold degeneracy is broken near the interface, causing the CBM to split into two states. Fig.~\ref{fig:fig3} shows the summed and averaged atomic layer periods in the thickness direction for the two CBM states. For clarity, averaging was performed over the atomic layer period in the thickness direction. The results showed that there was a difference in the wave function spread between the N-doped and undoped states. When the external electric field was 0, the wave function peaked on the left side for both 0.6 nm and 2.4 nm nitrogen doping. This indicates that nitrogen doping lowers the energy levels of the CBM in the doped region, which suggests that the electron density in the said region is higher.

Interestingly, when the electric field was 5 MV/cm, there was no significant difference in the wave function spread between the nitrogen-doped depths of 0.6 nm and 2.3 nm compared to the undoped case. Conversely, when nitrogen was doped in the 3.7 nm region, the wave function spread was observed to move away from the interface. A similar trend can be observed when the electric field is 10 MV/cm. This suggests that when doped in the 3.7 nm region, the peak position of the wave function is farther away from the interface, resulting in less scattering and improved mobility. It should be noted that the experimental results show that the Hall mobility of SiC-MOSFETs does not improve (and may even decrease) with or without nitridation \cite{mobility4, hall1}. The present calculation results show that there is no difference in the peak value and position of the wave function when the nitrogen doping area is narrow (doping depth of 0.6 nm and 2.4 nm); therefore, the Hall mobility is not improved, which is consistent with the calculation results. The actual diffusion of N into the SiC region owing to the nitridation process may be less than 5$\times$10$^{19}$ cm$^{-3}$ within approximately 2.4 nm. Conversely, considering that the wave function changes significantly when the nitrogen diffusion region reaches 3.7 nm, the mobility of the electrons themselves (Hall mobility) may be improved if nitrogen can be doped at a high density of only a few nm above the interface.

\begin{figure}
\includegraphics[keepaspectratio, scale=0.34]{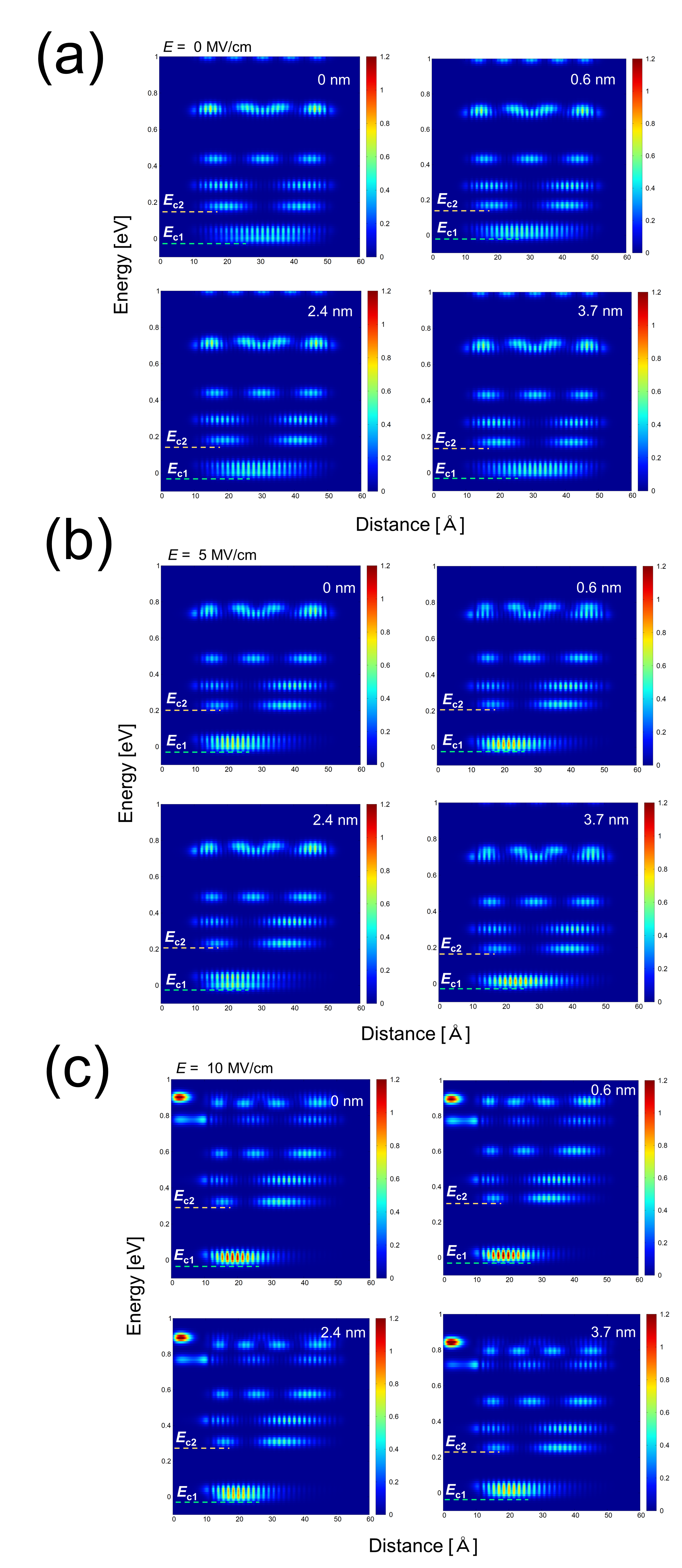}
\caption{\label{fig:fig4}Calculated LDOS of the conduction bands when the externally applied electric field is (a) 0, (b) 5, and (c) 10 MV/cm. Doping depths of N are 0, 0.6, 2.4, and 3.7 nm. LDOS is indicated by the color code. The vertical axis is the energy and the energy level of the CBM is set to 0 in each figure.}
\end{figure}
 
 Fig.~\ref{fig:fig4} shows the LDOS near $E_{\rm C}$. Regardless of the presence or absence of nitrogen doping, the energy difference between the first ($E_{\rm C1}$) and second sub-band ($E_{\rm C2}$) increases as the applied electric field increases. For example, in the case of no nitrogen doping, $E_{\rm C2}-E_{\rm C1}$ is 0.17 eV when the applied electric field is 0, 0.23 eV when the applied electric field is 5 MV/cm, and 0.30 eV when the applied electric field is 10 MV/cm. The increase in $E_{\rm C2}-E_{\rm C1}$ with increasing applied voltage is thought to be due to the steeper gradient of the triangular confinement potential at the interface and the larger quantum confinement effect. Therefore, by comparing the energy difference between $E_{\rm C1}$ and $E_{\rm C2}$, the strength of the quantum confinement effect can be determined.

\begin{figure*}
\includegraphics{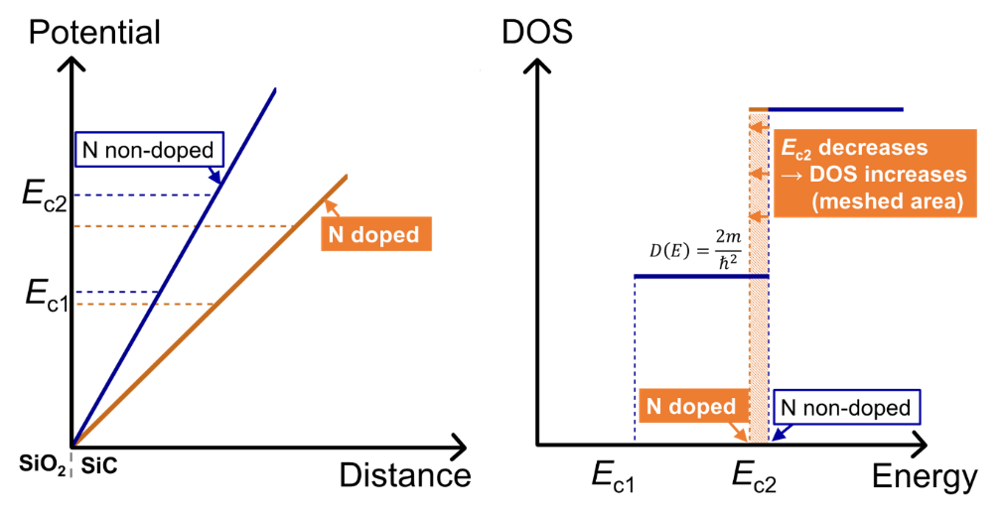}
\caption{\label{fig:fig5}Energy shift of $E_{\rm C2}$ when the surface is densely doped with nitrogen. (a) Change in triangular potential with respect to $E_{\rm C1}$. (b) Change in DOS due to the energy shift of $E_{\rm C2}$.}
\end{figure*}

To emphasize, when the nitrogen doping depth is up to 2.4 nm, $E_{\rm C2}-E_{\rm C1}$ is almost unchanged compared to the undoped case, but when the doping depth is extended to 3.7 nm, the difference between the first and second sub-band is found to be narrower when an electric field is applied. For example, when the electric field strength is 5 MV/cm, the energy difference between $E_{\rm C1}$ and $E_{\rm C2}$ without doping is 0.23 eV, and that with a nitrogen doping depth of 3.7 nm is 0.19 eV. This indicates that the quantum confinement effect is smaller when nitrogen is doped in the 3.7 nm region. This is thought to be due to the fact that the amount of band bending is reduced by nitrogen doping compared to the undoped case. This is consistent with the fact that the potential flattens as the doping level increases, as shown in Fig. ~\ref{fig:fig2}. A smaller quantum confinement effect is thought to result in a lower energy position of $E_{\rm C2}$ with respect to $E_{\rm C1}$, thus increasing the probability of electron occupancy in $E_{\rm C2}$ (Fig. ~\ref{fig:fig5}(a)). Fig.~\ref{fig:fig5}(b) shows a schematic of the density of states (DOS), where the number of free carriers in the conduction bands increases as the DOS increases for the same $E_{\rm f}$. It has not yet been determined whether the origin of the SiC-MOS interface level is localized or in a tail state based on the conduction-band edge of two-dimensional DOS \cite{matsushita1, matsushita2, matsushita3, yoshioka1, yoshioka2}. Therefore, the free electron density relative to the total electron density (free electrons and electrons trapped in the interface defects) is expected to be larger when N is doped. This indicates that $D_{\rm it}$ may be reduced.

The results at this stage suggest that when approximately 2.4 nm of highly doped N is introduced on the surface, there is almost no change in both free carrier density and mobility, but when 3.7 nm or more is introduced, free carrier density increases (apparent $D_{\rm it}$ decreases) and mobility improves. Conversely, a high density of nitrogen at the interface leads to an increase in the free carrier density (decrease in apparent $D_{\rm it}$) and an increase in the Hall mobility.
It is widely known that the formation of a thin n-type layer (doping concentration $<$ 1$\times$10$^{17}$ cm$^{-3}$, thickness of $\sim$100–300 nm) immediately below the gate oxide is effective in improving peak channel mobility (up to 80–-170 cm$^{2}/$Vs) \cite{Ueno, harada3, Okuno, Kimoto}.
However, when the gate voltage is increased to the bias required for practical operation (oxide field $>$ 2 MV/cm$^{-1}$), mitigation of the surface electric field owing to the presence of the n-type doped layer is weakened and the mobility decreases sharply. 
In this study, compared with the undoped case, the wave function is further away from the interface and $E_{\rm C2}-E_{\rm C1}$ is smaller in the n-doped case, suggesting that quantum confinement is weakened even under an extremely high electric field of 10 MV/cm.
Considering the above results, if a few nm of a heavily doped n-type region ($>$ 1$\times$ 10$^{19}$) is formed on the SiC surface, high channel mobility may be obtained even when a high electric field is applied.
An important perspective for industrial applications is that MOSFETs should be designed to be normally in the off-state.
In this study, it has not yet been investigated whether normal off-state operation can be achieved for SiC-MOSFETs with a heavily doped ultrathin n-type layer below the gate oxide film.
Therefore, further studies are required for practical applications.


\section{\label{sec:Conclusions}Conclusions}
In this study, first-principles calculations were performed to investigate the behavior of electron carriers at the SiC/SiO$_2$ interface when a few nm of the SiC interface was nitridated. It was found that a relatively wide nitridation region of approximately 4 nm from the interface with a nitrogen doping density of 5$\times$ 10$^{19}$ cm$^{-3}$ has a significant effect on the LDOS and wave function. When an electric field is applied, the quantum confinement effect decreases when the nitridation region is 4 nm, which is thought to increase the electron density induced in the inversion layer. As for the wave function, even when an electric field was applied, the peak shifted significantly toward the direction that pulled the electrons away from the interface. This is thought to reduce the effect of electron scattering at the interface and improve electron mobility. Therefore, the insertion of a few nm ($\sim$4 nm) of a highly doped nitrogen layer ($\sim$ 5 $\times$ 10$^{19}$ cm$^{-3}$) on the interface of n-channel SiC-MOSFETs is expected to improve both the electron density and electron mobility. 

\begin{acknowledgments}
This work was supported by MEXT as a ``Program for Promoting Researches on the Supercomputer Fugaku'' (JPMXP1020200205) and by JSPS KAKENHI as a ``Grant-in-Aid for Scientific Research(A)'' (Grant Number 21H04553, 20K05352, 20H00340, 22H01517, 22K18292). The computation in this work was performed using the supercomputer Fugaku provided by the RIKEN Center for Computational Science/Supercomputer Center at the Institute for Solid State Physics in the University of Tokyo.

\end{acknowledgments}

\appendix

\section{Validity of VCA}
The validity of VCA in the present study was examined. As part of the validity study, we evaluated the difference between the vacuum level and SiC potential in 4H-SiC slabs when N was substituted by the supercell method and when N was doped by VCA.

The 4H-SiC supercell (11$\bar{2}$0) was used in this calculation, as described in the main text.
The supercell used in the slab calculation had a structure with five atomic layers in the in-plane (3 × 1) thickness direction (a = 16.1 \AA, b = 34.0 \AA, c = 10.1 \AA). The vacuum region was set to 19 \AA, and the number of atoms was 264.

The pseudopotentials of the atoms used for the undoped and nitrogen-substituted cases were those of the undoped Si, C, and N cases.
In the case of VCA calculations, the pseudopotentials for the case of no Si doping were used, assuming 1\% N doping with VCA for the C atoms only. This corresponded to a doping density of 5 $\times$ 10$^{20}$ cm$^{-3}$.
The supercell method was used for the case of N in the slab. This corresponded to a doping density of 4 $\times$ 10$^{20}$ cm$^{-3}$. Although the doping densities were slightly different between the VCA and nitrogen substitution cases, we determined that this was not an issue in validating VCA.

The potential distribution in the SiC slab was averaged in the atomic period thickness direction to determine the average SiC potential. The amplitude of the averaged potential was found to be 2 meV at most. The potential at the slab edge ($E_{\rm vac}$) was determined relative to the SiC potential averaged in the slab. This was performed for the undoped, nitrogen doping in the supercell, and VCA cases.

The results show that in the VCA case, the difference between the average potential and $E_{\rm vac}$ is 0.29 eV higher than that of the undoped case.
However, in the case of nitrogen doping, the difference was 0.22 eV higher than that of the undoped case. This result confirms that the potential change is reproduced well by VCA; therefore, the effect of nitrogen doping was incorporated into this study by employing VCA.

\nocite{*}

\bibliography{apssamp}

\end{document}